\documentclass[onecol]{svjour2}
\usepackage{amsmath}
\usepackage{color}

\begin{document}
\newcommand{\tr}{{\rm Tr}}
\newcommand{\E}{{\mathcal E}}

\title{Equivalent definitions of the quantum nonadiabatic entropy production}
\author{Jordan M.\ Horowitz \and Takahiro Sagawa}

\institute{J.\ M.\ Horowitz \at
Department of Physics, University of Massachusetts at Boston, Boston, MA 02125, USA \\
\email{Jordan.Horowitz@umb.edu}
\and
T. Sagawa\at
Department of Basic Science, The University of Tokyo, 3-8-1 Komaba, Meguro-ku, Tokyo 153-8902, Japan
}

\date{\today}

\maketitle

\begin{abstract}
The nonadiabatic entropy production is a useful tool for the thermodynamic analysis of continuously dissipating, nonequilibrium steady states.
For open quantum systems, two seemingly distinct definitions for the nonadiabatic entropy production have appeared in the literature, one based on the quantum relative entropy and the other based on quantum trajectories.
We show that these two formulations are equivalent.
Furthermore, this equivalence leads us to a proof of the monotonicity of the quantum relative entropy under a special class of completely-positive, trace-preserving quantum maps, which circumvents difficulties associated with the noncommuntative structure of operators.

\keywords{Quantum nonequilibrium thermodynamics \and Nonadiabatic entropy production \and Quantum relative entropy monotonicity}
\end{abstract}

\section{Introduction}\label{sec:intro}

A diversity of physical systems are prevented from relaxing to thermodynamic equilibrium by nonconservative forces or nonequilibrium boundary conditions.
Instead, they are  maintained in a nonequilibrium steady state characterized by continuously dissipating currents~\cite{Parrondo2002,Seifert2012,Esposito2009}.
As thermodynamic systems, the second law of thermodynamics requires their entropy production rate to always be positive, ${\dot S}_{\rm tot}\ge 0$.
Unfortunately, this inequality is uninformative for driven transitions between distinct nonequilibrium steady states, because the constant dissipation required to sustain a steady state masks any subtle effects due to the external driving.
 This raises the question as to what extent can thermodynamics be usefully applied to far-from-equilibrium processes?
In a effort to construct a useful far-from-equilibrium formalism, several approaches have been introduced to isolate the effect of the external driving by subtracting off the constant steady state dissipation from the total dissipation~\cite{Hatano2001,Trepagnier2004,Speck2005b,Esposito2010,Ge2010,Komatsu2008,Sagawa2011,Maes2012,Bertini2013}. 
The approach we address in this paper accomplishes this task through a refinement of the second law, where the entropy production rate is divided into two separate, positive contributions~\cite{Hatano2001,Trepagnier2004,Speck2005b,Esposito2010,Ge2010}:
\begin{equation}
{\dot S}_{\rm tot}={\dot S}_{\rm a}+{\dot S}_{\rm na}.
\end{equation}
The adiabatic piece ${\dot S}_{\rm a}\ge 0$ captures the baseline entropy production due to the dissipation in the steady state.
The nonadiabatic entropy production rate ${\dot S}_{\rm na}\ge 0$ quantifies the additional dissipation required to drive the system between steady states~\cite{Hatano2001}, thereby isolating the effect of the driving.
Of particular importance is the fact that ${\dot S}_{\rm na}$ is always positive: this positivity restricts the possible set of processes that can occur -- in much the way the second law of thermodynamics does -- which is key for its utility as a theoretical tool.

For classical dynamics, the nonadiabatic entropy production rate is well characterized~\cite{Hatano2001,Esposito2010,Ge2010,Sagawa2011,Mandal2013,Spinney2012}.
By contrast,  ${\dot S}_{\rm na}$  has been defined for open quantum systems in two seemingly disparate ways.
The first, due to Yukawa \cite{Yukawa2001}, is in terms of a quantum relative entropy, and positivity follows from  the monotonicity of the quantum relative entropy under completely-positive, trace-preserving (CPTP) quantum maps~\cite{Nielsen2000,Hayashi2006,Sagawa2012}.
The second definition is formulated in terms of the statistics of quantum trajectories~\cite{Horowitz2013d}.
Positivity is demonstrated using well-known classical probability inequalities, such as Jensen's inequality~\cite{Cover}.
 Having two different possible definitions of ${\dot S}_{\rm na}$ suggests that the quantum nonadiabatic entropy production may not play a role in a thermodynamic formalism for far-from-equilibrium quantum processes.
However, in this article we reconcile these two approaches, thereby taking a step in establishing a unified formalism.

After reviewing both quantum definitions of ${\dot S}_{\rm na}$ in Secs.~\ref{sec:setup}, \ref{sec:relEnt}, and \ref{sec:traj}, we demonstrate in Sec.~\ref{sec:equiv} that they are in fact equivalent.
At first sight this equivalence is surprising; while both definitions imply the positivity of ${\dot S}_{\rm na}$, the relative entropy formulation relies on the full mathematical machinery of noncommunative operator algebras, while the trajectory formulation uses simpler classical reasoning.
We resolve this incongruence in Sec.~\ref{sec:proof}, where we prove, using classical arguments, the monotonicity of a special class of quantum relative entropies that have the interpretation of a nonadiabatic entropy production for a CPTP map.
This analysis offers a distinct perspective on the monotonicity of the quantum relative entropy, and its sister problem of the subadditivity of the von Neumann entropy, which has been one of the major problems in statistical mechanics since the seminal work of Lieb~\cite{Lieb1973,Lieb1973c,Lieb1973b}.
Beyond the field of statistical mechanics, the monotonicity has attracted attention the last two decades in quantum information theory~\cite{Nielsen2000,Hayashi2006,Sagawa2012}, where many important theorems (e.g., the Holevo bound) can  be directly obtained from the monotonicity, and the strong subadditivity has been useful in applications of the AdS/CFT correspondence in string theory~\cite{Headrick2007}.


\section{Setup}\label{sec:setup}

To begin our discussion, let us first introduce the basic ingredients common to the two definitions of ${\dot S}_{\rm na}$.
We have in mind a finite quantum system weakly coupled to its surroundings, which we model as a collection of ideal thermodynamic reservoirs that can exchange energy and/or matter.
We call this collection of reservoirs the \emph{environment}.
To control the system we have access to a collection of external parameters $\lambda$, which we vary with time $t$ according to a prescribed protocol $\lambda_t$ in order  to either manipulate the system Hamiltonian $H(\lambda)$ or alter the intensive variables of the reservoirs, such as their temperatures or chemical potentials.
For such open quantum systems, the dynamics of the density matrix $\rho_t$ is well described by the master equation~\cite{Yukawa2001,Horowitz2013d,Nechita2009,Horowitz2012,Breuer,Chetrite2012}
\begin{equation}\label{eq:master}
\dot\rho_t=-i[H(\lambda_t),\rho_t]+\sum_k {\mathcal D}[L_k(\lambda_t)] \rho_t \equiv {\mathcal L}(\lambda_t)\rho_t,
\end{equation}
where ${\mathcal D}[c]\rho=c\rho c^\dag-(1/2)c^\dag c \rho-(1/2)\rho c^\dag c$.
The Lindblad or jump operators $L_k$ represent different possible transitions that can be observed in the environment.
For example, the jump operators of a  thermal reservoir induce transitions where  a quanta of energy (or heat) is exchanged.
We further assume, that when we fix $\lambda$, the system will relax to a unique, positive-definite, steady state with density matrix $\pi_\lambda$, given as the solution of ${\mathcal L}(\lambda)\pi_\lambda=0$.
For clarity of notation, the steady-state density matrix evaluated along the parameter protocol $\lambda_t$ at time $t$ will simply be denoted as $\pi_t\equiv\pi_{\lambda_t}$.

To facilitate our subsequent analysis, we also briefly recall a useful splitting of ${\dot S}_{\rm na}$ akin to the standard division of the total entropy production into the system and environment pieces~\cite{Esposito2010b,VandenBroeck2010}:
\begin{equation}\label{eq:split}
{\dot S}_{\rm na}(t)={\dot S}(t)+{\dot S}_{\rm ex}(t).
\end{equation}
Here, the first term is the rate of change of the system's von Neumann entropy $S(t)=-\tr[\rho_t\ln\rho_t]$ (with Botzmann's constant set to $k=1$), which due to probability conservation ($\tr[\rho_t]=1$) is expressible simply as 
\begin{equation}\label{eq:S}
{\dot S}(t)=-\tr[\dot\rho_t\ln\rho_t],
\end{equation}
whose derivation we recall in Appendix~\ref{sec:appendix2}.
The second contribution ${\dot S}_{\rm ex}(t)$ is called the excess entropy production rate and represents the additional flow of entropy into the environment over and above that required to maintain the steady state.
For isothermal processes at temperature $T$, this is sometimes reexpressed in terms of the excess heat current into the reservoir, ${\dot Q}_{\rm ex}=T{\dot S}_{\rm ex}$~\cite{Seifert2012,Hatano2001,Speck2005b,Ge2010,Yukawa2001}, which in the absence of nonconservative, external forces reduces to the usual heat flow.
The precise mathematical expression for ${\dot S}_{\rm ex}$ differs in the two setups, which we now discuss.


\section{Relative entropy formulation}\label{sec:relEnt}

In this section, we review a definition of the nonadiabatic entropy production rate in terms of the quantum relative entropy.

For two operators $\chi$ and $\phi$, their \emph{quantum relative entropy} is~\cite{Nielsen2000,Hayashi2006,Sagawa2012}
\begin{equation}
D(\chi \| \phi) = \tr [ \chi \ln \chi] - \tr [ \chi \ln  \phi].
\end{equation}
In terms of $D$, Yukawa~\cite{Yukawa2001} has identified the nonadiabatic entropy production rate as
\begin{equation}\label{eq:NArelent}
\dot{S}_{\rm na}(t) =  - \frac{\partial}{\partial s}D(\rho_s||\pi_t) \Big|_{s=t} = \dot S(t) + \dot S_{\rm ex}(t), 
\end{equation}
where ${\dot S}$ is given in (\ref{eq:S}), and the excess entropy production rate is 
\begin{equation}\label{eq:Exrelent}
\dot S_{\rm ex} (t)= \tr[\dot{\rho}_t\ln\pi_t].
\end{equation}

The positivity of $\dot{S}_{\rm na}$ is a consequence of the monotonicity of the quantum relative entropy, that is $D$ decreases under the action of a CPTP map.
To make this statement precise, consider a CPTP map with Kraus representation~\cite{Nielsen2000,Hayashi2006,Sagawa2012}
\begin{equation}\label{eq:E}
\mathcal E (\rho) = \sum_k M_k \rho M_k^\dagger,
\end{equation}
where the trace preserving property is imposed by
\begin{equation}\label{eq:TP}
\sum_k M_k^\dagger M_k = I,
\end{equation}
with $I$ the identity operator.
Then the quantum relative entropy monotonically decreases for any $\chi$ and $\phi$ under the action of any CPTP map~\cite{Nielsen2000,Hayashi2006,Sagawa2012}:
\begin{equation}
 D(\chi||\phi)\ge D (\mathcal E (\chi ) \| \mathcal E (\phi )).
\label{monotonicity1}
\end{equation}
In light of (\ref{eq:NArelent}), we are interested in the special situation where $\phi$ is the invariant state of the map $\pi=\E(\pi)$, in which case~(\ref{monotonicity1}) reduces to 
\begin{equation}
D(\chi \|  \pi) \geq D (\mathcal E (\chi ) \| \pi ).
\label{monotonicity2}
\end{equation}
Proving the monotonicity of the quantum relative entropy is more challenging than its classical counterpart.
The difficulty lies in the noncommutative operator structure, which can be treated using either operator monotonicity or convexity. 
Historically, the  monotonicity of  the quantum relative entropy  was proved by Lindblad and Uhlmann~\cite{Lindblad1974,Lindblad1975,Uhlmann1977} on the basis of the strong subadditivity of the von Neumann entropy~\cite{Lieb1973,Lieb1973c,Lieb1973b}.  Later,  proofs by Petz and Nielsen were put forward that did not invoke strong subadditivity, but instead relied on operator convexity~\cite{Nielsen2005,Petz1986,Petz2003}.

To see how monotonicity implies positivity of ${\dot S}_{\rm na}$, we rewrite (\ref{eq:NArelent}) as
\begin{equation}
\dot{S}_{\rm na}(t) = \lim_{\Delta t \to 0} \frac{D(\rho_t \|  \pi_t ) - D (\rho_{t+\Delta t} \| \pi_t) }{\Delta t}.
\end{equation}
Since the solution of the master equation is a CPTP map that induces the infinitesimal evolution ${\mathcal E}_{t,t+\Delta t}:\rho_t \to  \rho_{t+\Delta t}$~\cite{Breuer} and preserves the steady state ${\mathcal E}_{t,t+\Delta t}(\pi_t)=\pi_t$, we observe from inequality (\ref{monotonicity2}) that
\begin{equation}
D(\rho_t \|  \pi_t ) - D (\rho_{t+\Delta t} \| \pi_t ) \geq 0,
\end{equation}
and
\begin{equation}
\dot{S}_{\rm na}(t) \geq 0.
\end{equation}
In this sense, the positivity of the nonadiabatic entropy production rate is a straightforward consequence of the monotonicity of the quantum relative entropy.
We stress that the monotonicity is a very general property that holds for all CPTP maps.


\section{Quantum trajectory formulation}\label{sec:traj}

As an alternative to the relative entropy formulation of the previous section, we now review an approach based on quantum trajectories introduced in  \cite{Horowitz2013d}.

The quantum trajectory description of open quantum systems is based on a continuous monitoring of environmental transitions~\cite{Breuer,Brun2002}.
As we observed in Sec.~\ref{sec:setup}, the different jump operators $L_k$ in (\ref{eq:master}) represent the different possible transitions in the environment.
In this way, whenever an environmental transition is observed, the backaction on the system is induced by the corresponding $L_k$.
The quantum trajectory is the random evolution of the system conditioned on the environmental transitions~\cite{Breuer,Brun2002,Jacobs2006,Wiseman1994}.
We denote by  $\varrho_t$ the density matrix at time $t$ \emph{conditioned} on all the transitions up to time $t$, and its evolution is given by the stochastic master equation~\cite{Wiseman1994}
\begin{equation}\label{eq:StochMaster}
\begin{split}
d\varrho_t=&-idt\big\{{\mathcal H}(\lambda_t)\varrho_t-\varrho_t{\mathcal H}^\dag(\lambda_t)-\tr\left[{\mathcal H}(\lambda_t)\varrho_t-\varrho_t{\mathcal H}^\dag(\lambda_t)\right]\varrho_t\big\} \\
&+\sum_k dN_k(t) \left[\frac{{\mathcal J}_k(\lambda_t)\varrho_t}{\tr[{\mathcal J}_k(\lambda_t)\varrho_t]}-\varrho_t\right],
\end{split}
\end{equation}
where ${\mathcal H}(\lambda)=H(\lambda)-\frac{i}{2}\sum_kL_k^\dag(\lambda)L_k(\lambda)$ is an effective non-hermitian Hamiltonian and ${\mathcal J}_k(\lambda)\varrho = L_k(\lambda)\varrho L^\dag_k(\lambda)$ is the jump superoperator.
Here, the $dN_k(t)$ are Poisson increments that are $1$ when the $k$ transition is observed, and $0$ otherwise.
They are defined by the condition $dN_k(t)^2=dN_k(t)$ -- which enforces that they be only $0$ or $1$ -- and by their classical expectation
\begin{equation}\label{eq:EdN}
E[dN_k(t)]=\tr[{\mathcal J}_k(\lambda_t)\rho_t]dt=\tr[ L^\dag_k(\lambda_t)L_k(\lambda_t)\rho_t] dt.
\end{equation}
Its Poisson statistics are evident here, since its expectation is of order $dt$.
From $\varrho_t$, we can recover the unconditioned density matrix by taking the classical expectation, $\rho_t=E[\varrho_t]$.

In this formulation, the proof of the positivity of ${\dot S}_{\rm na}$ rests on the master equation satisfying a number of additional requirements~\cite{Horowitz2013d}.
When these requirements are satisfied, we say the description is \emph{thermodynamically consistent}.
First, we assume that if a transition $k$ is possible, then the reverse transition, denoted ${\tilde k}$, is also possible.
Thus, the jump operators come in pairs, $L_k$ and $L_{\tilde k}$.
Furthermore, these jump operators are related through a local detailed balance relation~\cite{Horowitz2013d}
\begin{equation}
L_{k}(\lambda)=L_{\tilde k}^\dag (\lambda) e^{\Delta s_k(\lambda)/2},
\end{equation}
where $\Delta s_k(\lambda)$ is the entropy flow into the environment during a $k$ jump.
For a thermal reservoir, $\Delta s_k=q_k/T$, where $q_k$ is the heat flow into the reservoir during the $k$ transition.
At their heart, these properties are a consequence of the time-reversal symmetry of the underlying microscopic dynamics, which demands that any physical process can occur in reverse.
Next, we assume that the jump operators be time-reversal symmetric, that is they commute with the time-reversal operator $\Theta$~\cite{Sakurai}:
\begin{equation}
L_k(\lambda)=\Theta L_k(\lambda)\Theta.
\end{equation}
To introduce our last requirement, we must first recall that for any master equation (\ref{eq:master}), its representation in terms of jump operators is not unique~\cite{Breuer}.
However, we can single out a \emph{privileged representation} with respect to the unique, positive-definite $\pi_\lambda$, by assuming that the system's dual dynamics is also CPTP~\cite{Fagnola2007} (or equivalently that the generator ${\mathcal L}(\lambda)$ commutes with the modular operator $\sigma_{-i}^\lambda(\rho)=\pi_\lambda\rho\pi_\lambda^{-1}$: ${\mathcal L}(\lambda)\sigma_{-i}^\lambda=\sigma_{-i}^\lambda{\mathcal L}(\lambda)$).
In this case, the privileged representation is defined by the properties~\cite{Fagnola2007} 
\begin{align}\label{eq:priv1}
\pi_\lambda L_k(\lambda)\pi_\lambda^{-1}&=\varpi_k(\lambda) L_k(\lambda)\\
\label{eq:priv2}
[H(\lambda),\pi_\lambda]&=[\sum_kL_k^\dag(\lambda)L_k(\lambda),\pi_\lambda]=0,
\end{align}
where the $\varpi_k$ are positive constants that verify $\varpi_k=\varpi_{\tilde k}^{-1}$ and are given as ratios of the eigenvalues of $\pi_\lambda$.
Then, our last requirement for a thermodynamically consistent description is that (\ref{eq:master}) is expressed in its privileged representation.
Alternatively, we may assume that the jump operators $L_k$ in (\ref{eq:master}) satisfy (\ref{eq:priv1}) and (\ref{eq:priv2}).
While it may seem like a strong restriction to require (\ref{eq:priv1}) and (\ref{eq:priv2}) to be valid, many systems of physical interest seem to satisfy this constraint~\cite{Horowitz2013d}.

In  \cite{Horowitz2013d}, ${\dot S}_{\rm na}$ is defined as a function of  the quantum trajectory.
For our purposes here, it is worthwhile to take a slightly different, but equivalent point of view.
Instead, the nonadiabatic entropy production rate is expressed in terms of the stochastic increment
\begin{equation}\label{eq:NAincr}
ds_{\rm na}(t)=ds(t)+ds_{\rm ex}(t).
\end{equation}
Here, $ds(t)$ is the stochastic differential of the stochastic system entropy $s(t)=-\tr[\varrho_t\ln\rho_t]$, which is defined so that its classical expectation is the von Neumann entropy $E[s(t)]=S(t)$~\cite{Horowitz2013d,Horowitz2012}.
The stochastic excess entropy production rate is
\begin{equation}
ds_{\rm ex}(t)=\sum_k dN_k(t)\ln \varpi_k(\lambda_t).
\end{equation}
We see that $ds_{\rm ex}$ depends only on $dN_k(t)$ and not directly on $\varrho_t$; it changes only when there are jumps.
This is a signature that the excess entropy production rate is a flow and not a state function like the entropy.

To compare with the relative entropy formalism, we need the average rate 
\begin{equation}\label{eq:NAtraj}
{\dot S}_{\rm na}(t)\equiv\frac{E[ds_{\rm na}(t)]}{dt}={\dot S}(t)+{\dot S}_{\rm ex}(t),
\end{equation}
where ${\dot S}$ is in (\ref{eq:S}) and
\begin{equation}\label{eq:EXtraj}
{\dot S}_{\rm ex}(t)=\frac{E[ds_{\rm ex}(t)]}{dt}=\sum_k\tr[L_k^\dag(\lambda_t)L_k(\lambda_t)\rho_t]\ln\varpi_k(\lambda_t),
\end{equation}
by virtue of (\ref{eq:EdN}).
We have used the same notation to represent the entropy production rates in the relative entropy and trajectory formulations.
In the following section, we will justify this choice by demonstrating that they are the same.

Before continuing, let us briefly comment on the proof of the positivity of ${\dot S}_{\rm na}$ in this setup.
The stochastic increment $ds_{\rm na}(t)$ in (\ref{eq:NAincr}) is a classical stochastic process, even though its statistics originate from quantum fluctuations.
Thus, it is amenable to classical probability and information theory analysis~\cite{Cover}.
To be specific, let us consider the stochastic nonadiabatic entropy production during the time interval $t=0$ to $\tau$, $\Delta s_{\rm na}(\tau)=\int_0^\tau ds_{\rm na}(t)$.
Then the treatment in \cite{Horowitz2013d}  implies that 
\begin{equation}
E\left[ e^{-\Delta s_{\rm na}}\right] = 1,
\end{equation}
which can be verified using standard arguments for proving and analyzing fluctuation theorems~\cite{Harris2007}.
It then follows from Jensen's inequality~\cite{Cover} for real convex functions that 
\begin{equation}
\Delta S_{\rm na}=E[\Delta s_{\rm na}]\ge 0.
\end{equation}
We see that once the trajectory formulation is set up, the positivity of the nonadiabatic entropy production follows readily.


\section{Equivalence of formulations}\label{sec:equiv}

Having seen that there are two distinct expressions for the nonadiabatic entropy production rate, we demonstrate in this section that the two definitions are the same for thermodynamically consistent descriptions.

To begin, first note that by construction the system contribution ${\dot S}$ in the two formulations [(\ref{eq:NArelent}) and (\ref{eq:NAtraj})] are the same.
Thus, we turn to the excess contributions.
We demonstrate their equality by rewriting the relative entropy definition in (\ref{eq:Exrelent}) using the defining properties of the privileged representation in (\ref{eq:priv1}) and (\ref{eq:priv2}), as follows.
Starting with (\ref{eq:Exrelent}), we insert the master equation (\ref{eq:master}), and rearrange using the commutation relations of the privileged representation (\ref{eq:priv2}):
\begin{equation}
{\dot S}_{\rm ex}(t)=\sum_k\tr\left[\left(L^\dag_k(\lambda_t)\ln(\pi_t)L_k(\lambda_t)-L^\dag_k(\lambda_t) L_k(\lambda_t) \ln \pi_t\right)\rho_t\right].
\end{equation}
Next, we utilize an identity for the privileged representation, 
\begin{equation}\label{eq:lnIdent}
\ln(\pi_\lambda) L_k(\lambda)=L_k(\lambda)\ln(\varpi_k(\lambda)\pi_\lambda),
\end{equation}
which we derive in Appendix \ref{sec:appendix} as a straightforward consequence of (\ref{eq:priv1}), to arrive at
\begin{align}
{\dot S}_{\rm ex}(t)&=\sum_k\tr\left[\left(L^\dag_k(\lambda_t)L_k(\lambda_t)\ln(\varpi_k(\lambda_t)\pi_t)-L^\dag_k(\lambda_t) L_k(\lambda_t) \ln\pi_t\right)\rho_t\right] \\
&=\sum_k\tr[L_k^\dag(\lambda_t)L_k(\lambda_t)\rho_t]\ln\varpi_k(\lambda_t),
\end{align}
which reproduces the trajectory definition of (\ref{eq:EXtraj}).


\section{Classical proof of the monotonicity of the quantum relative entropy}\label{sec:proof}

In the preceding sections, we have shown that two different definitions of the nonadiabatic entropy production rate are equivalent.
Both definitions imply that ${\dot S}_{\rm na}\ge 0$, but by different means.
The quantum relative entropy approach relies on the monotonicity of the quantum relative entropy; by contrast, the quantum trajectory formulation uses classical probability inequalities.
In this section, we provide insight into this difference.
We prove the monotonicity of a special class of quantum relative entropies using well-known convexity properties for functions of real variables, without relying on operator techniques.

We go back to the setup in Sec.~\ref{sec:relEnt}, with the CPTP map $\E(\rho)=\sum_k M_k\rho M_k^\dag$ on a finite dimensional Hilbert space introduced in (\ref{eq:E}), with
invariant state $\pi=\E(\pi)$.
As before, we assume $\E$ has a privileged representation with Kraus operators that satisfy
\begin{equation}\label{eq:mu}
\pi M_k \pi^{-1}=\mu_k M_k,
\end{equation}
with $\mu_k>0$.
We can guarantee the existence of the privileged representation by assuming that the dual map $\tilde\E(\rho)=\pi^{-1}\E^\dag(\pi\rho)$ is CPTP (or by making the equivalent assumption that $\E$ commutes with the modular operator $\sigma_{-i}(\rho)=\pi \rho \pi^{-1}$)~\cite{Fagnola2007}.
Then (\ref{eq:mu}) follows directly from Lemma 4.1 of \cite{Fagnola2007}.

We will now prove that the relative entropy $D(\rho||\pi)$ decreases under the action of $\E$ for all $\rho$, that is
\begin{align}
\Delta D&=D(\E(\rho)||\pi)-D(\rho||\pi)\le 0,
\end{align}
by rewriting $\Delta D$ as a function of real numbers.

As a first step, we introduce the diagonal bases of 
\begin{equation}\label{eq:W}
\begin{split}
\rho&=\sum_m P_m |p_m\rangle\langle p_m| \\
\E(\rho)&=\sum_\alpha P^\prime_\alpha|e_\alpha\rangle\langle e_\alpha|=\sum_{\alpha,m,k} w^k_{\alpha,m} P_m |e_\alpha\rangle\langle e_\alpha|,
\end{split}
\end{equation}
where $P_m,P^\prime_\alpha>0$ and $\sum_m P_m=\sum_\alpha P^\prime_\alpha=1$ with $w^k_{\alpha,m}=|\langle e_\alpha|M_k|p_m\rangle|^2$ forming the elements of a stochastic matrix $W_{\alpha,m}=\sum_k w^k_{\alpha,m}$ that verifies $\sum_\alpha W_{\alpha,m}=1$~\footnote{$W$ is a stochastic matrix, since $\E$ preserves the trace (\ref{eq:TP}).}.
Next, we divide $\Delta D$ into two smaller pieces that are easier to analyze: 
\begin{align}
\Delta D&=\big(\tr[\E(\rho)\ln\E(\rho)]-\tr[\rho\ln\rho]\big)-\big(\tr[\E(\rho)\ln\pi]-\tr[\rho\ln\pi]\big) \\
&\equiv d_{\rm S}+d_{\rm ex},
\end{align}
where the labeling is  by analogy to the nonadiabatic entropy production rate with system contribution $d_{\rm S}$ and excess contribution $d_{\rm ex}$.
Let us first simplify $d_{\rm S}$ by applying the decomposition in (\ref{eq:W}):
\begin{align}
d_{\rm S}&=\sum_{\alpha}P_\alpha^\prime\ln P_{\alpha}^\prime - \sum_{m}P_m\ln P_m \\
\label{eq:ds}
&=\sum_{\alpha,m,k}w^k_{\alpha,m}P_m\ln P_\alpha^\prime-\sum_{\alpha,m,k}w^k_{\alpha,m}P_m\ln P_m.
\end{align}
For $d_{\rm ex}$, we insert the definition of $\E$ in (\ref{eq:E}) as well as (\ref{eq:TP}),
\begin{align}
d_{\rm ex}&=-\tr[\sum_k M_k\rho M_k^\dag \ln \pi]+\tr[\rho\sum_k M^\dag_k M_k \ln\pi]\\
&=-\sum_k\tr\left[\big(\ln(\pi) M_k-M_k\ln \pi\big)\rho M_k^\dag\right].
\end{align}
Assuming the privileged representation (\ref{eq:mu}), we again have the identity [cf.~(\ref{eq:lnIdent})]
\begin{equation}
\ln(\pi)M_k=M_k\ln(\mu_k\pi),
\end{equation}
which implies that
\begin{equation}\label{eq:dex}
d_{\rm ex}=-\sum_k\tr[M_k\rho M_k^\dag]\ln\mu_k=-\sum_{\alpha,m,k}w^k_{\alpha,m}P_m\ln\mu_k.
\end{equation}
Finally, summing (\ref{eq:ds}) and (\ref{eq:dex}):
\begin{equation}
\Delta D=d_{\rm S}+d_{\rm ex}=\sum_{\alpha,m,k}w^k_{\alpha,m}P_m\ln\frac{P^\prime_\alpha}{\mu_k P_m},
\end{equation}
which is now expressed completely in terms of real numbers.
Thus, all the complications of the noncommutativity of operators have been removed.
This lends insight into why ${\dot S}_{\rm na}$ in the trajectory formulation could be expressed as a classical expectation;
$\Delta D$ for this setup is essentially classical in  nature despite being an operator expression.

We complete the proof using the convexity of the logarithm $\ln x\le x-1$,
\begin{equation}
\Delta D\le \sum_{\alpha,m,k}w^k_{\alpha,m}P_m\left(\frac{P^\prime_\alpha}{\mu_k P_m}-1\right)=0,
\end{equation} 
where the equality follows from the conservation of probability $\sum_m P_m=\sum_\alpha P_\alpha^\prime=1$ and by recognizing that the definition of $\mu_k$ in (\ref{eq:mu}) implies that 
\begin{equation}
\sum_{k,m}\frac{1}{\mu_k}w^k_{\alpha,m}=\sum_k \frac{1}{\mu_k}\langle e_\alpha|M_kM_k^\dag|e_\alpha\rangle=\sum_k\langle e_\alpha|\pi^{-1}M_k\pi M_k^\dag |e_\alpha\rangle= 1.
\end{equation}


\section{Conclusion}

The nonadiabatic entropy production can be a useful theoretical tool for gaining insight into the thermodynamics of far-from-equilbrium systems.
 We have shown that two seemingly disparate definitions of ${\dot S}_{\rm na}$ for open quantum systems are in fact the same. 
 This provides us with the possibility of approaching the thermodynamics of open quantum systems using either the ensemble picture (relative entropy formulation) or in terms of individual fluctuating trajectories, just as in classical stochastic thermodynamics~\cite{Seifert2012,Sekimoto}.
 It further lends weight to the observation that we have identified an appropriate quantum analog of the classical nonadiabatic entropy production.
 However, these definitions are not the only possibilities.  
  A third approach has been suggested by Chetrite and Mallick~\cite{Chetrite2012}, and later by Liu~\cite{Liu2012}, that utilizes a quantum extension of the Feynman-Kac formula for quantum Markov semi-groups.
While there is no general connection between their approach and those considered here, a recent work by Liu has pointed to such a possibility~\cite{Liu2013}.
 
 We also investigated the monotonicity of the quantum relative entropy under the action of a class of CPTP maps.
The relative entropy is an operator expression.
However, we have seen the relative entropy with respect to the map's invariant state, can be expressed totally in terms of real numbers.
In this sense, $\Delta D$ is a classical quantity, despite its definition in terms of operators.
This seems to imply that the existence of a CPTP dual map restricts quite severely the structure of a quantum map, making it nearly classical in a certain sense.
 

\appendix

\section{Time derivative of von Neumann entropy}\label{sec:appendix2}

The derivation of (\ref{eq:S}) follows by first taking the derivative of the von Neumann entropy $S(t)=-\tr[\rho_t\ln\rho_t]$,
\begin{equation}
\dot S(t) = -  \tr [ \dot \rho_t \ln \rho_t ] -   \tr [\rho_t \frac{d}{dt}\ln \rho_t ].
\end{equation}
Clearly, (\ref{eq:S}) is true if the second term is zero.
To demonstrate this, we decompose $\rho_t$ in its time-dependent  eigenbasis  $\rho_t = \sum_k P_t^k | k_t \rangle \langle k_t |$, where $\sum_k P_t^k = 1$ and  $\{ | k_t \rangle \}$ form an orthonormal basis.  We then have 
\begin{align}
\tr [\rho_t \frac{d}{dt}\ln \rho_t ] &= \tr [\rho_t \sum_k \frac{\dot P_t^k }{P_t^k} | k_t \rangle \langle k_t |] + \tr [\rho_t \sum_k \ln P_t^k \frac{d}{dt}| k_t \rangle \langle k_t |]  \\
&= \sum_k \dot P_t^k + \sum_k  P_t^k \ln P_t^k \frac{d}{dt}\tr [| k_t \rangle \langle k_t |] \\ 
&= 0.
\end{align}

\section{Privileged representation identity}\label{sec:appendix}

To verify (\ref{eq:lnIdent}), we first expand the logarithm in its power series about $I$ and then exploit (\ref{eq:priv1}) for the privileged $L_k$ as
\begin{align}
\ln(\pi_\lambda) L_k(\lambda)&=\sum_{n=1}^\infty \frac{(-1)^{n+1}}{n}(\pi_\lambda-I)^n L_k(\lambda) \\
&=L_k(\lambda)\sum_{n=1}^\infty\frac{(-1)^{n+1}}{n}(\varpi_k(\lambda)\pi_\lambda-I)^n \\
&=L_k(\lambda)\ln(\varpi_k(\lambda)\pi_\lambda).
\end{align}
It is worth noting that this argument will hold for other functions besides the logarithm as long as they have a power series expansion.

\begin{acknowledgements}
We are grateful to Franco Fagnola for providing the proof of the existence of a privileged representation for CPTP maps.
JMH is supported by ARO MURI grant W911NF-11-1-0268 and TS by JSPS KAKENHI Grant Nos. 25800217 and 22340114.
\end{acknowledgements}

\end{document}